%% file: AMNPPRS.tex
\documentclass[12pt,twoside]{article}

\usepackage{epsf,cite} 
\usepackage{colordvi}

\input{SA_macros.tex}

\begin{document}
\pagestyle{myheadings} 
\markboth
{\it SUSY Long-Lived Massive Particles...}
{\it SUSY Long-Lived Massive Particles...}

\begin{titlepage}
\thispagestyle{empty} 

\null~\vspace{-2.0cm} 

\begin{flushright} 
hep-ph/0012192     
\hfill  
CERN-TH/2000-349
\end{flushright} 

\hrule\hfill

\vspace{-0.1cm}

\hrule\hfill

\vspace{0.2cm} 
                                        
\begin{center}
\BrickRed{{\Huge \bf SUSY Long-Lived Massive Particles:} \\
{\Huge {\bf Detection and Physics at the LHC}}}\\

\vspace{1.5cm}

{\large {\bf S.~Ambrosanio}~$^{a,1}$,  
        {\bf B.~Mele}~$^{b,c}$,
        {\bf A.~Nisati}~$^{b,c}$,
        {\bf S.~Petrarca}~$^{c,b}$,
        {\bf G.~Polesello}~$^{d}$,
        {\bf A.~Rimoldi}~$^{d,e}$,
        {\bf G.~Salvini}~$^{c,b}$} \\
~\\
$^a$ CERN -- {\it Theory Division,
     CH-1211 Geneva 23, Switzerland}\\
~\\
$^b$ INFN -- {\it Sezione di Roma I , c/o Dipartimento di Fisica, 
     Universit\`a ``La Sapienza'',}\\
     {\it p.~le Aldo Moro 2, I-00185 Roma, Italy}\\
~\\
$^c$ {\it Dipartimento di Fisica, 
     Universit\`a ``La Sapienza'',}\\
     {\it p.~le Aldo Moro 2, I-00185 Roma, Italy}\\
~\\
$^d$ INFN -- {\it Sezione di Pavia e Dip.~Fis.~Nucl.~Teor.,
     Universit\`a di Pavia,}\\
     {\it via Bassi 6, I-27100 Pavia, Italy}\\
~\\
$^e$ CERN -- {\it EP Division,
     CH-1211 Geneva 23, Switzerland}\\
\end{center}

\vspace*{\fill}  

\centerline{\bf Abstract}
~\\
\noindent
{\small 
We draw a possible scenario for the observation of massive long-lived 
charged particles at the LHC detector ATLAS.
The required flexibility of the detector triggers and of the identification 
and reconstruction systems are discussed.
As an example, we focus on the measurement of the mass and lifetime  
of long-lived charged sleptons predicted in the framework
of supersymmetric models with gauge-mediated supersymmetry (SUSY) breaking. 
In this case, the next-to-lightest SUSY particle can be the 
light scalar partner of the tau lepton ($\tauu$), possibly decaying slowly 
into a gravitino. A wide region of the SUSY parameters space was explored. 
The accessible range and precision on the measurement of the SUSY breaking 
scale parameter $\sqrt{F}$ achievable with a counting method are assessed.
}

\vspace*{\fill}  

\centerline{Submitted for publication to {\it Rendiconti dei Lincei}}

\vspace*{\fill}

\noindent 
\parbox{0.4\textwidth}{\hrule\hfill} \\ 
{\small $^1$ Address after 1--Jan--2001: \\ 
\null~~~Banca di Roma, Direzione Generale, Linea Finanza, 
viale U.~Tupini 180, I-00144 Roma, Italy
}  

\end{titlepage}

\setcounter{page}{0} 

~\null
\thispagestyle{empty}
\newpage 

\thispagestyle{plain}

\section{Introduction} 
\label{sec:intro} 

\noindent 
The Large Hadron Collider (LHC) at the European Laboratory for 
Particle Physics (CERN) in Geneva has an enormous and unique
potential of discovering new particles beyond the Standard Model 
(SM) over a large mass range around the TeV scale. \\
In this article, we consider an interesting class of exotic particles 
that are characterised by a few model independent common features:

\begin{description} 
\item[a)] they have large masses and are produced with $\beta < 1$
      in a non negligible fraction of events at the LHC;
\item[b)] they are stable or quasi-stable, i.e. their lifetime is larger
 than  $10^{-7}$ s, so that they  decay  far from the collision point,
 possibly outside the
 detector; 
\item[c)] their electric charge is either an integer or a fractional
 multiple of the proton charge.
\end{description}

Actually, quite a few theories beyond the SM foresee 
charged particles, both strongly and electroweakly interacting, with
large masses and, although many exotic particles are
unstable, often the lowest (or the next-to-lowest) 
lying state may be stable or quasi-stable~\cite{Errede,kaori}. 
In the following,
we call these states Massive Semi-stable Exotic Particles (MSEP's).\\
In the first part of this paper, we will briefly recall
the difficulties which a large apparatus like ATLAS could meet
while trying to detect MSEP's.
In the second part, we show a concrete example of 
the measurements (involving detection, identification and 
track reconstruction) of MSEP's coming from supersymmetric 
models~\cite{StevePrimer}  with gauge-mediated supersymmetry
 breaking (GMSB)~\cite{oldGMSB,newGMSB,GR-GMSB}.
In this scheme, the supersymmetry (SUSY)~\cite{StevePrimer} 
breaking occurs at relatively low energy scales, and it is mediated 
mainly by gauge interactions. The automatic suppression of the SUSY 
contributions to flavour-changing neutral currents and CP-violating 
processes is naturally fulfilled. Furthermore, in the simplest versions 
of GMSB, the Minimal Supersymmetric Standard Model (MSSM) spectrum 
and other observables depend on just five parameters, usually
chosen to be~\cite{GMSBmodels1,GMSBmodels2,AKM-LEP2,AB-LC}: 
the overall messenger scale $M_{\rm mess}$, the so-called messenger index 
$N_{\rm mess}$, the universal soft SUSY breaking scale felt by the 
low-energy sector $\Lambda$, the ratio of the vacuum expectation 
values of the two Higgs doublets $\tan\beta$, and sign($\mu$), which  
is the ambiguity left for the SUSY higgsino mass after imposing the 
conditions for a correct electroweak symmetry breaking.

In the following, we will consider GMSB models where the r\^ole of MSEP
is taken by the stau (or all charged sleptons), which is the 
next-to-lightest SUSY particle (NLSP), decaying into gravitinos.
These scenarios are very promising at the LHC, providing signatures 
of semi-stable charged tracks coming from massive sleptons, 
therefore  with  $\beta$ often significantly smaller than 1. 
In particular, we perform
our simulations at the ATLAS muon detector, whose
excellent time resolution~\cite{TDR} allows a precision measurement of
the slepton time of flight, and hence of the slepton velocity.
We show that the event can be  recognised with high efficiency
by the calorimetric trigger requiring the usual SUSY signature of 
$P_{T}^{\rm miss} +$jets, where the hadronic jets, originated by the 
decay of heavy squarks/gluinos along with MSEP's, must have 
the transverse momentum $P_T>50$~GeV 
and the transverse momentum imbalance $P_{T}^{\rm miss}$ is calculated only 
from the energy deposition in the hadronic calorimeter.
It is remarkable that, in this framework,
the event rate is very high and the 
background contamination due to SM channels can be reduced
to  a completely negligible level. 
Then, we conclude stressing the fact
that the measurements of mass (using the MSEP momentum obtained by
the muon system) and lifetime (obtained by a counting method) are sufficient
to determine the SUSY breaking scale $\sqrt{F}$ at   
a level of precision of a few 10$\%$'s.

Although this topic did not receive much attention up to now,
as we will discuss, there is a sound theoretical basis to emphasise 
the physics and experimental search for MSEP's.

\section{The Experimental Challenge} 
\label{sec:challenge} 
The experimental signatures of heavy long-lived charged 
particles at a hadron collider have  been studied both 
in the framework of GMSB and in more general 
scenarios~\cite{leandro,drtata,femoroi,marthom}.
The two main observables one can use to separate
these particles from muons are the high specific ionisation
and the time of flight in the detector.
As a matter of fact,  a MSEP behaves like
a massive muon  with a velocity sometimes considerably lower than $c$.
The energy loss of a MSEP in matter has been carefully studied 
in Ref.~\cite{Tata} (see also Ref.~\cite{Errede}), where it was shown that,
for energy loss for ionisation when $\beta > {1\over2}$ and for particles 
with $M > $100 GeV/$c^2$, the range in iron exceeds several metres
and increases with $M$. Moreover, for strongly interacting MSEP's
the penetration length in matter does not change significantly
for $\beta \leq 0.7$ and $M= $100 GeV/${\rm c}^2$~\cite{Tata}. 
At the same time, a remarkable property of MSEP's is that, because of 
their low $\beta$ value, they can produce anomalous energy loss for 
ionisation in matter that can be used as a distinctive feature for their 
identification.

\begin{figure}[t] 
\centerline{
\epsfxsize = \textwidth
\epsffile{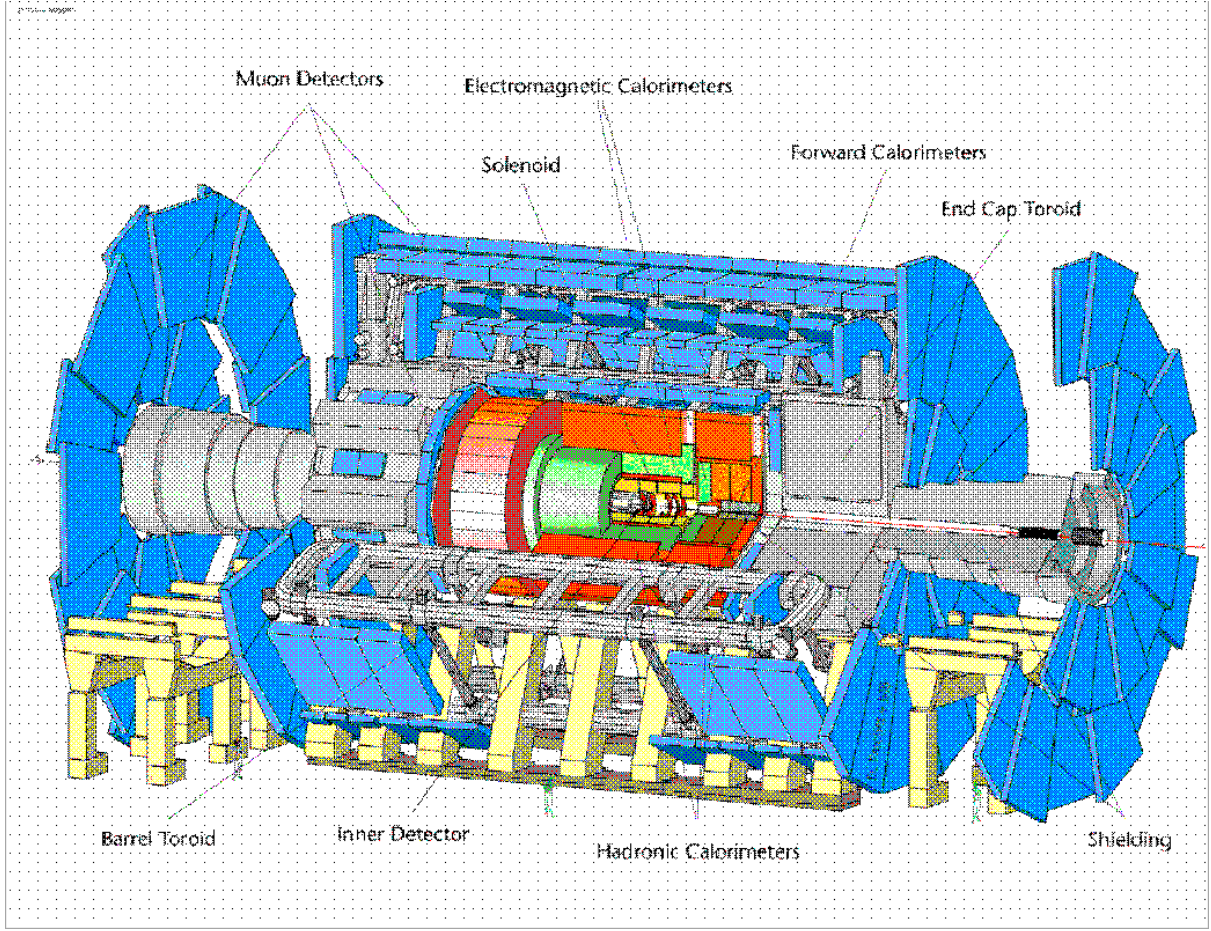}
}
\caption{\sl Isometric view of the ATLAS detector
with  the description of the different functional parts.
}
\label{fig:ATLAS-iso}
\end{figure}

\begin{figure}[t] 
\centerline{
\epsfxsize = \textwidth 
\epsffile{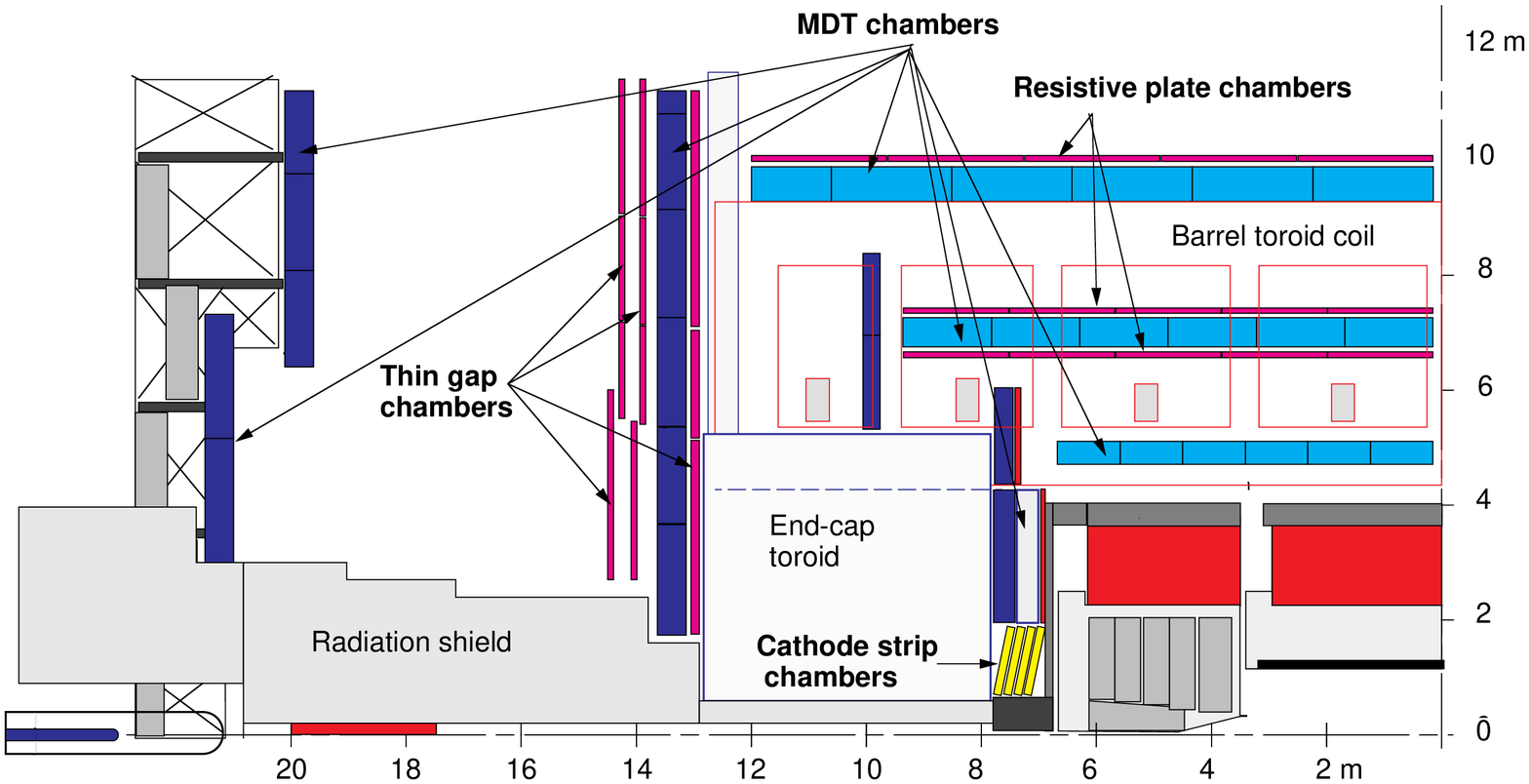}
}
\caption{\sl  Side view of the ATLAS detector
with  the description of the different functional parts.
}
\label{fig:ATLAS-side}
\end{figure}

Nevertheless, there are some drawbacks associated to low-velocity
MSEP's. In fact, the detector synchronisation is  tuned for particles 
travelling across the apparatus with the velocity of light, hence,
due to the large geometrical dimensions of the typical LHC apparatus,
all the data acquisition system should be carefully tuned in order to 
guarantee the trigger and the event reconstruction of a MSEP. 
The trigger system of the ATLAS experiment is described in detail
in Ref.~\cite{TDR} (see Figs.~\ref{fig:ATLAS-iso} and~\ref{fig:ATLAS-side}
for  schematic views of the ATLAS apparatus). Three levels of trigger 
are envisaged, which will reduce the initial input rate of 40 MHz for the 
first-level trigger progressively down to a rate of $\sim$ 100 Hz. 
This is the maximum input rate of the data storage system, and corresponds 
to an acquisition of $\sim$ 100 MB/s. The first level comes
exclusively from the hardware, and the information from the muon  detectors
and from the calorimeters are treated separately.  The second level refines
the first level by connecting the information from different detectors.
Finally, the third level (also called {\it event filter}) applies the
full off-line reconstruction algorithms to the data.

The track identification of a MSEP escaping the hadronic calorimeter in the 
high $p_T$ muon region is the natural task of the ATLAS specialised structure
of muon precision chambers, composed by MDT's (Monitored Drift Tubes), which 
are dedicated to measure the trajectory of penetrating particles and to
determine, with a good precision, the position of a charged particle. 

We recall that in the central region an air-core toroidal superconducting 
magnet generates a magnetic field of about 0.5 T to allow the measurement of 
the particle momentum. The precision chambers are composed by two multilayers,
and each of them is composed of three (four in the inner station) drift tube 
layers. A typical track goes through about twenty MDT's, before escaping from 
the apparatus. A single MDT is a cylindrical drift tube of 3 cm diameter. 
It is basically filled with Argon + CO$_2$ at 3 atmosphere absolute pressure. 
It has a typical spatial resolution $\sigma = 80$ $ \mu$m and a maximum drift 
time of 700 ns. The resulting spatial resolution for the single 
particle is about 30 $\mu$m~\cite{frontend}. 
Once an event has been accepted by the first level trigger, all the data 
coming from the MDT's for the following $700$ ns are labelled with the same 
bunch crossing number and are extracted from the pipe-line memories. 
The track reconstruction for a MSEP will be done by taking into account in 
the hit analysis the delay along the track  due to $\beta < 1$.

In our case, we will assume to trigger the MSEP's 
event by the hadronic calorimeter which can be easily activated by the 
multiple high $P_T$ jets produced by the decay of squarks/gluinos with 
masses larger than 500~GeV together with MSEP's with masses of the 
order of 100~GeV. The classical SUSY signature of  
$P_{T}^{\rm miss} +$jets, where $P_{T}^{\rm miss}$ is 
calculated only from the energy deposit in 
the calorimeter, neglecting the NLSP's and the muons, is in fact enough
to pass the requests of the first level $P_{T}^{\rm miss}$~ trigger
of a jet with $P_T>50$~GeV and $P_{T}^{\rm miss} > 50$~GeV. 
A drawback of this purely calorimetric approach is the fact that processes
with low hadronic activity, such as direct slepton production and direct
electroweak gaugino production are not selected.  

The measurements of the time of flight for MSEP's are made possible by the 
timing precision ($\ltap 1$~ns) and the size of the ATLAS muon spectrometer.

In the barrel part of the detector ($|\eta|<1$), the precision
muon system consists of three multilayers of precision drift tubes
immersed in a toroidal air-core magnetic field.
The three measuring stations are located at distances of approximately 
5, 7.5 and 10 m from the interaction point. A particle crossing a drift 
chamber ionises the chamber gas along its path, and the
electrons produced by the ionisation drift to the anode wire
under the influence of an electric field. The particle position
is calculated from the  measurement of the drift time of the
ionisation electrons to the anode wire. In order to perform this calculation,
a starting time $t_0$ for counting the drift time is needed, corresponding
to the time of flight of the particle from the production
point to the measuring station.
For a particle travelling  approximately at the speed of light, as a muon,
the $t_0$'s for the measuring stations are parameters of the detector geometry
and of the response of the front-end electronics~\cite{leandro1}.
For a heavy particle, the $t_0$ is a free parameter, function of the 
$\beta$ (= v/c) of the particle.
It was demonstrated with a full simulation of the ATLAS muon 
detector~\cite{gpar} that the $\beta$ of a particle can be measured by 
adjusting the $t_0$ for each station in such a way to minimise the 
$\chi^2$ of the reconstructed muon track.

The  particle $\beta$ can be measured with a resolution approximately 
parameterised as $\sigma(\beta)/\beta^2=0.028$.
The resolution on the transverse momentum measurement for heavy
particles is found to be comparable to the one expected for muons.

Besides the strong signature of the delay needed to reconstruct the track, 
a MSEP is identified by the energy loss for ionisation that can be much 
higher than that of a muon at minimum.
This signature is a powerful physical quantity that gives
a clear MSEP's identification.

As discussed in Ref.~\cite{leandro}, already at $\beta \sim 0.6$ 
the energy loss is around twice the minimum, and of course it 
increases rapidly for a lower value of $\beta$.
A complete discussion of the ionisation measurement in the 
ATLAS detector is a very complicated problem which can not be addressed
in this work. Here, we want to illustrate only in a preliminary way the 
possible r\^ole of the hadronic calorimeter and of the MDT's.
We believe that the calorimeter can bring a significant contribution
to the identification of MSEP's. The ATLAS hadron calorimeter can measure
the muon energy deposit with an accuracy of about 25\%, in the last 
compartment, and $\sim$ 20\%, using the full depth. The contamination induced
by the minimum bias event pile-up is expected to be of the order of 1\% 
or lower at full luminosity.

In the case of the MDT detector, the measurement of total charge based on a
single MDT appears poor because of the background contamination 
at full LHC luminosity.
Notwithstanding, measurement of charge performed by a statistical procedure
involving the complete set of $\sim$ 20 tubes fired by the particle can 
provide a significantly improved measurement.
It must be considered that the MDT's front-end electronics is optimised to 
provide a charge integration only for a small fraction of the total drift 
time ($\sim 30$~ns). However, even with this limitation, it could be possible
that again the statistical combination of all the tubes belonging to the
track can provide a good ionisation measurement.

Of course, this brief discussion does not exhaust all the possibilities 
offered by the ATLAS apparatus.
For example, it was shown in Ref.~\cite{TDR} that the ionisation 
energy loss measurement in the Transition Radiation Tracker 
can be used to achieve $\pi/K$
separation and, in some cases, the electromagnetic calorimeter 
can provide the charge measurements.

\section{Supersymmetric Scenario}
\label{sec:susy} 

\noindent
The phenomenology of GMSB (and more in general of any theory with low-energy
SUSY breaking) is characterised by the presence of a very light gravitino 
$\G$~\cite{Fayet},
\be 
m_{3/2} \equiv m_{\G} = \frac{F}{\sqrt{3}M'_P},  
\label{eq:Gmass}
\ee
\noindent
where $\sqrt{F}$ is the fundamental scale of SUSY breaking.
If we assume the typical value $\sqrt{F}=100$ TeV, 
 and $M'_P = 2.44 \times 10^{18}$ GeV for  the reduced 
Planck mass, we get $m_{3/2}=2.37$ eV.
Hence,  $\G$ is always the lightest SUSY particle (LSP) in these theories. 
If $R$-parity is assumed to be conserved, any produced MSSM particle will 
finally decay into the gravitino. Depending on $\sqrt{F}$, the interactions 
of the gravitino, although much weaker than gauge and Yukawa interactions, 
can still be strong enough to be of relevance for collider physics. 
As a result, in most cases the last step of any SUSY decay chain is 
the decay of the next-to-lightest SUSY particle (NLSP), which can  
occur outside or inside a typical detector (even close to the interaction 
point). The pattern of the resulting spectacular signatures is determined 
by the identity of the NLSP and its lifetime before decaying into  $\G$,

\be 
\frac{c \tau_{\rm NLSP}}{\rm cm} \simeq \frac{1}{100 {\cal B}}
\left(\frac{\sqrt{F}}{100 \; {\rm TeV}}\right)^4 
\left(\frac{m_{\rm NLSP}} {100 \; {\rm GeV}}\right)^{-5},
\label{eq:NLSPtau}
\ee 

\noindent
where ${\cal B}$ is a number of order unity depending on the nature
of the NLSP.
The identity of the NLSP [or, to be more precise, the identity of the 
sparticle(s) having a large branching ratio (BR) for decaying into the 
gravitino and the relevant SM partner] determines four main scenarios 
giving rise to qualitatively different phenomenology: a) the Neutralino-NLSP
Scenario; b) the Stau-NLSP Scenario; c) the Slepton co-NLSP Scenario; 
d) the Neutralino-Stau co-NLSP Scenario~\cite{AKM-LEP2}. In this paper,
we will be concerned with Scenarios b) and c) only.  

The two main parameters affecting the experimental measurement at the
LHC of the slepton NLSP properties are the slepton mass and momentum 
distribution. Indeed, at a hadron collider most of the NLSP's come
from squark and gluino production, followed by cascade decays. 
Thus, the momentum distribution is in general a function of the whole 
MSSM spectrum. However, one can approximately assume that most of the 
information on the NLSP momentum distribution is provided by the squark 
mass scale $m_{\tilde q}$ only (in the stau NLSP scenario or slepton 
co-NLSP scenarios of GMSB, one generally finds 
$m_{\tilde{g}} \gtap m_{\tilde q}$). 
To perform detailed simulations, we select a representative set of 
GMSB models generated by {\tt SUSYFIRE}~\cite{SUSYFIRE}. 
We limit ourselves to models with $m_{\rm NLSP} > 100$ GeV,
that cannot be excluded by direct searches at LEP/Tevatron, 
and $m_{\tilde q} < 2$ TeV, in order to yield 
an adequate event statistics after a three-year low-luminosity run 
(corresponding to 30 fb$^{-1}$) at the LHC.
Within these ranges, we choose eight extreme points (four in the stau 
NLSP scenario and four in the slepton co-NLSP scenario) allowed by GMSB
in the ($m_{\rm NLSP}$, $m_{\tilde q}$) plane, in order to cover the 
various possibilities. More details can be found in 
Refs.~\cite{AMPPR1,AMPPR2}. 
In Tab.~\ref{tab:tbg} (first five rows), we list the input GMSB parameters 
we used to generate these eight points.

\begin{table}[t]
\begin{center}
\begin{tabular}{|r||r|r|r|r|r|r|r|c|} \hline
Model & 1 & 2 & 3 & 4 & 5 & 6 & 7 & 8 \\ \hline\hline
$M_{\rm mess}$~($10^4$~TeV) & 1.79 & 5.28 & 0.0436 
& 0.0151& 3.88& 23.1 & 75.7& 0.0479 \\ \hline
$N_{\rm mess}$ &3&3&5&4&6&3&3&5\\ \hline
$\Lambda$~(TeV)&26.6&26.0&41.9&28.3&58.6&65.2&104& 71.9  \\ \hline
$\tan\beta$&7.22&2.28&53.7&1.27&41.9&1.83&8.54&3.27\\ \hline
sign($\mu$)& --& --& +& --& +& --& --& --\\ \hline\hline
$m_{\tilde\tau_1 = {\rm NLSP}}$ (GeV)&  100.1 &100.4 &101.0 &103.4 
&251.2 &245.3 &399.2 &302.9 \\ \hline
``NLSP'' Scenario &$\tauu$&$\tilde\ell$&$\tauu$&$\tilde\ell$&$\tauu$
&$\tilde\ell$ & $\tauu$&$\tilde\ell$ \\ \hline 
$m_{\tilde q}$ (GeV) &  577 &  563 &1190 &721 &1910 &1290 &2000 &1960 \\ \hline
$m_{\tilde g}$ (GeV)&  631 &617 &1480 &859 &2370 &1410 &2170 &2430 \\ \hline
$\sigma$ (pb)& 42& 50& 0.59& 10& 0.023& 0.36& 0.017& 0.022  \\ \hline\hline
SUSY events&452163&528420&7437 &147354     &365     &6535     &326 & 378  
\\ \hline 
BKGD:~$W$+Jets &9.6   &9.6   &9.5   &9.5   &2.4   &2.4   &1.0   &1.8 \\ \hline
BKGD:~$Z$+Jets~~&6.8  &6.9  &6.9  &7.1  &11.1  &11.0  &5.9  &8.7  \\ \hline
BKGD:~$\bar tt~~~$&5.3   &5.3   &5.3   &5.6   &6.2   &6.5   &3.4   &4.9
\\ \hline
BKGD:~QCD & 8.0   &8.0   &8.0   &7.4   &3.1   &3.1   &0.5   &2.0   \\ \hline
BKGD:~Total & 29.7&29.9& 29.9& 29.9& 22.8& 23.0& 10.8&17.4\\ \hline
\end{tabular}
\caption{\sl 
In rows 1--5, the input parameters of the eight sample GMSB models chosen 
for our study are reported. In rows 6--10, the main features of the models 
($\tilde{\ell} = \tilde{e}_R$, $\tilde{\mu}_R$, $\tauu$) are shown. 
The average squark mass is indicated by $m_{\tilde q}$ and $m_{\tilde g}$ 
is the gluino mass. In the last six rows, the comparison among the expected
number of events after the cuts described in Sect.~3.1 
and the major background sources are reported. 
The assumed integrated luminosity is 30~fb$^{-1}$, corresponding to a
three-year low-luminosity run at the LHC.
In the table, BKGD is a shorthand for ``background'', and $M_{\rm mess}$ and
$N_{\rm mess}$ are defined in the text 
(see Refs.~\cite{GR-GMSB,GMSBmodels1,GMSBmodels2,AKM-LEP2,AB-LC}). 
}
\label{tab:tbg}
\end{center}
\end{table}

\subsection{Event Selection and Slepton Mass Measurement} 
\label{sec:selection} 

\noindent
In order to select a clean sample of sleptons, we applied the
following requirements: 

\begin{itemize}
\item
at least a hadronic jet with $P_T>50$~GeV and a calorimetric \\
\mbox{$E_T^{\rm miss}>50$~GeV} (trigger requirement);
\item
at least one  slepton candidate satisfying the following cuts: 
\begin{itemize}
\item
$|\eta|<$2.4 to ensure that the particle is in the
acceptance of the muon trigger chamber, and therefore 
both coordinates can be measured;
\item
$\beta_{\rm meas}<0.91$, where $\beta_{\rm meas}$ is the $\beta$ of the
particle measured with the time of flight in the precision chambers;
\item
The $P_T$ of the slepton candidate, after the energy loss in the
calorimeters has been taken into account, must be larger than
10~GeV, to ensure that the particle traverses all  the muon stations.
\end{itemize}
\item
a cut $m_{\rm eff} > 800$ GeV, where $m_{\rm eff}$ is the total invariant mass
of the event constructed starting from the transverse momentum of the 
high $P_T$ jets and muons (or muon-like particles)~\cite{AMPPR2}.
\end{itemize} 

Considering an integrated luminosity of 30~fb$^{-1}$, 
a number of events ranging from a few hundred, for the models with
2~TeV squark-mass scale, to a few hundred thousand, for a 500~GeV
mass scale, survives these cuts. These events can be used to measure the 
NLSP properties. For the sake  of comparison, we recall that about 1500 
events for a light Higgs boson production are expected in the channel
$H\to \gamma\gamma$ in the same environment~\cite{TDR}.

In order to perform the mass measurement, the particle momentum is needed.
The precision chambers only provide a measurement of the momentum components  
transverse to the beam axis, so a measurement of the slepton pseudorapidity 
is necessary. This can be performed either by a match with a track
in the inner detector, or using the information from the muon trigger chambers.
The first option requires a detailed study of the matching procedure between 
detectors.
This study was performed for muons in Ref.~\cite{TDR}, but the results can 
not be transferred automatically to the case of  heavy particles for which
the effect of multiple scattering in crossing the calorimetric system
is much more severe. 

In the case of the trigger chambers, a limited time window around the beam 
crossing is read out, restricting the $\beta$ range for which the momentum 
can be measured.
We therefore evaluated the statistical precision achievable 
for the eight example models in two different $\beta$ intervals:
$0.6<\beta<0.91$ and $0.8<\beta<0.91$.

 Many more technical details about 
this analysis can be found in Refs.~\cite{AMPPR1,AMPPR2}. 

\subsection{Slepton Lifetime Measurement and $\sqrt{F}$}
\label{sec:lifetime}

\noindent 
The measurement of the slepton NLSP lifetime was performed by exploiting 
the fact that a couple of NLSP's is produced in each event. 
We adopted a method similar to the one used in Ref.~\cite{AB-LC} 
for the high energy $e^+e^-$ collider in the neutralino NLSP case.

This method consists in selecting the $N_1$ events where a slepton is detected 
through the time-of-flight measurement and then  counting how many times
($N_2$) a second slepton is observed. This information is sufficient
to determine the lifetime. 
Although in principle very simple, in practice this method requires 
an excellent control of all possible sources of inefficiency for 
detecting the second slepton.

\begin{figure}[t] 
\hspace{3.0cm}
\epsfxsize=0.7\textwidth
\epsffile{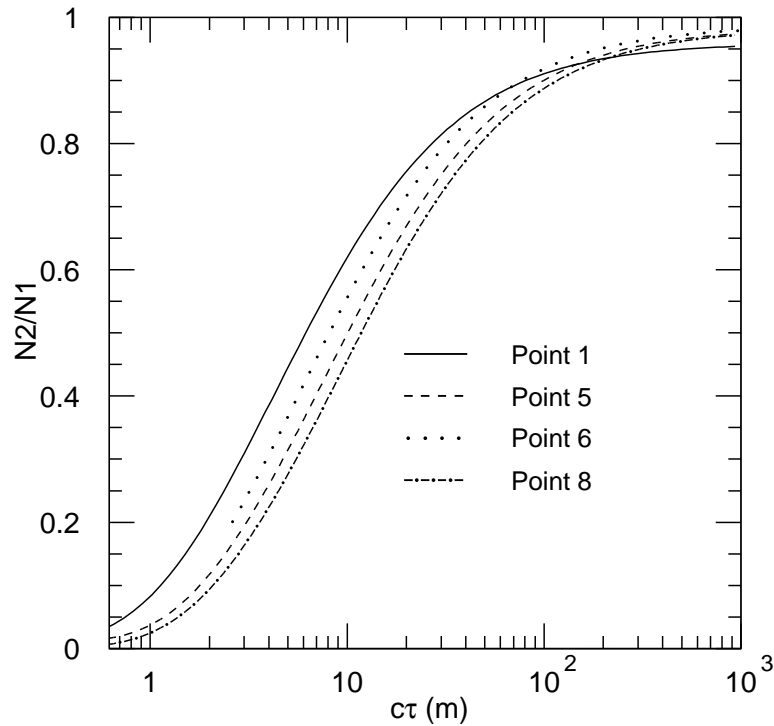}
\caption{\sl The ratio $R=N_2/N_1$ defined in the text as a function 
of the slepton lifetime $c\tau$. The curves corresponding to 
the model points 1, 5, 6, 8 are shown.} 
\label{fig:ratio}
\end{figure}

The first set of $N_1$ events is defined with the additional requirement that, 
for a given value of the slepton lifetime, at least one of the produced 
sleptons decays at a distance from the interaction vertex $>10$~m,
and is therefore reconstructed in the muon system.
For the events thus selected, we define $N_2$ as the number of events 
in the subsample where a second particle with transverse momentum 
$>10$~GeV is identified in the muon system. The search for the
second particle should be as inclusive as possible, in order to minimise
the corrections to the ratio
\be
R=\frac{N_2}{N_1},
\label{eq:ratio}
\ee
\noindent
which is a function of the slepton lifetime. Its dependence on the
NLSP lifetime $c\tau$ in meters is shown in Fig.~\ref{fig:ratio}
for four out of the eight sample models. 

The probability for a particle of mass $m$, momentum $p$ and proper 
lifetime $\tau$ to travel for a distance $L$ before decaying is
given by 
$P(L)=e^{-mL/pc\tau}$.
$N_2$ is therefore a function of the momentum distribution
of the slepton, which is determined by the details of the SUSY spectrum.
One needs therefore to be able to simulate the full SUSY cascade decays
in order to construct the $c\tau$--$R$ relationship. 

Relevant for the precision of the SUSY breaking scale  
measurement is the error on the measured $c\tau$. This can be 
extracted from the curves shown in Fig.~\ref{fig:ratio}.

\begin{figure}[t] 
\centerline{
\epsfxsize = 0.5\textwidth 
\epsffile{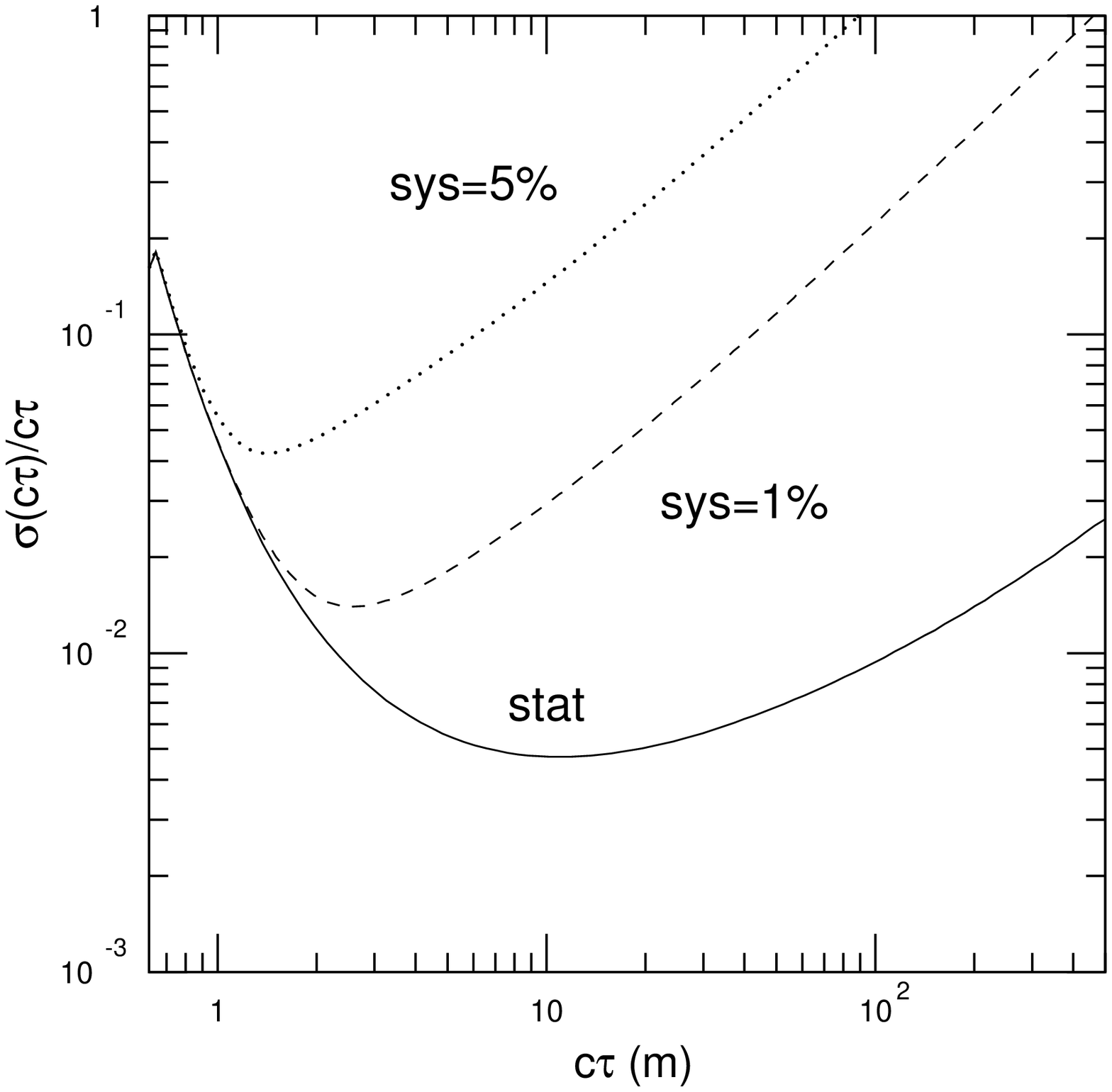}
\hfill
\epsfxsize = 0.5\textwidth 
\epsffile{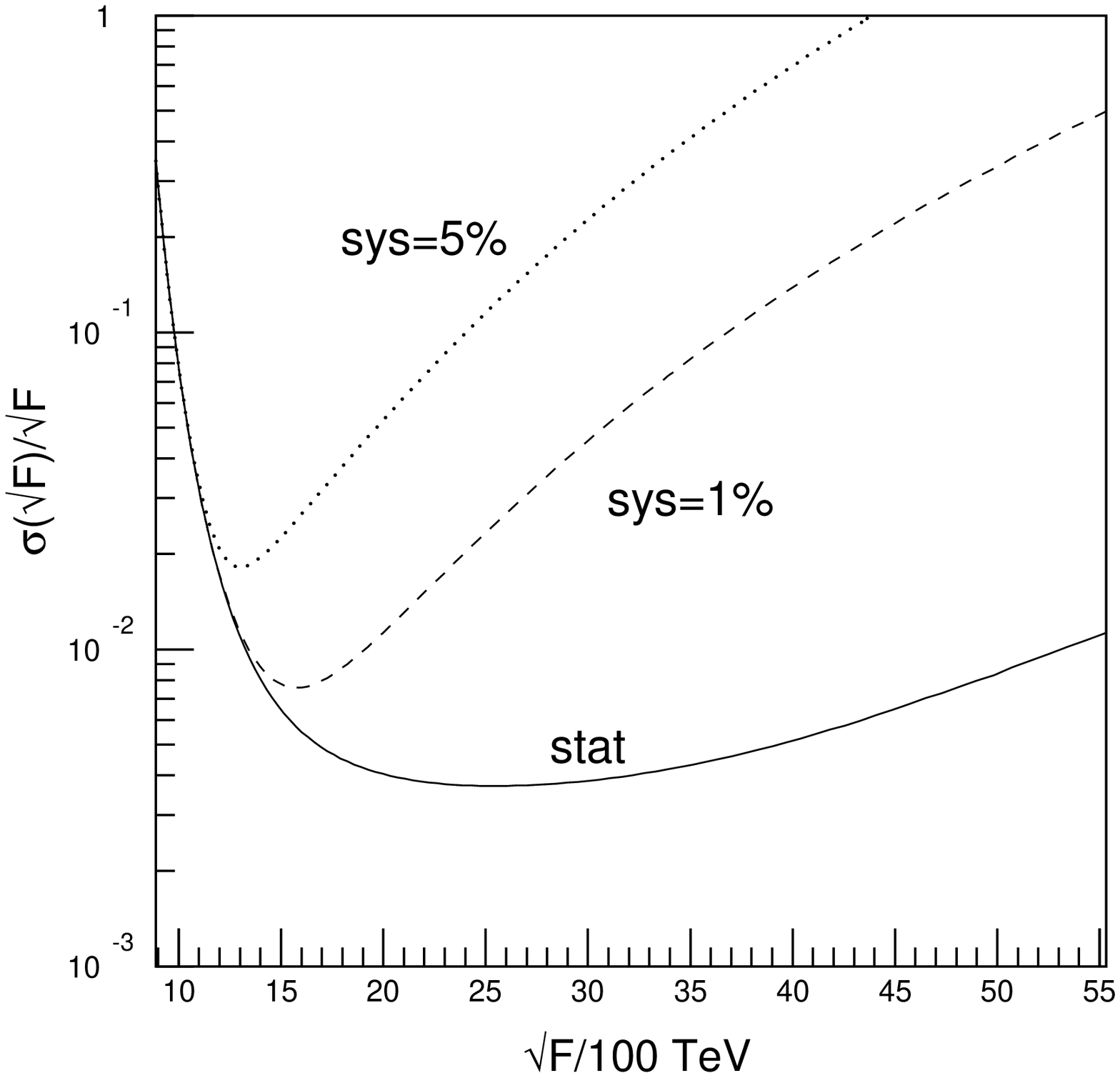}
}
\caption{\sl Fractional error on the measurements of the 
slepton lifetime $c\tau$ (left side) and the SUSY breaking scale
(right side), for model sample point 1 only.
We assume an integrated luminosity of 30~fb$^{-1}$. 
The curves are shown for three different assumptions on the 
fractional systematic error on the $R$ measurement: 
statistical error only (full line), 1\% systematic error (dashed line),
5\% systematic error (dotted line).
}
\label{fig:fractionalerror}
\end{figure}

The  precision calculated according to this formula is shown in 
Fig.~\ref{fig:fractionalerror} (left side), for model sample point 1 
and an integrated luminosity of 30~fb$^{-1}$. The full line in the plot 
is the error on $c\tau$ considering the statistical error on $R$ only. 
The available statistics is a function of the strongly interacting 
sparticle mass scale. 

We parameterise the systematic error as a term proportional to $R$, 
added in quadrature to the statistical error. We choose two values, 
$1\% R$ and $5\% R$, and propagate the error to the $c\tau$ measurement. 
The results are represented by the dashed and dotted lines in
Fig.~\ref{fig:fractionalerror} (left side).

For the models with squark mass scales up to 1200~GeV,
assuming a 1\% systematic error on the measured ratio, 
a precision better than 10\% on the $c\tau$ measurement 
can be obtained for lifetimes between 0.5--1~m and 50--80~m. 
If the systematic uncertainty grows up to 5\%, the 10\% precision
can only be achieved in the range 1--10~m. If the mass scale goes up to 2~TeV,
even considering a pure statistical error only, a 10\% precision
is not achievable. However a 20\% precision is possible  
over $c\tau$ ranges between 5 and 100~m, assuming a 1\% systematic error.

Using the measured values of $c\tau$ and the NLSP mass,
the SUSY breaking scale $\sqrt{F}$ can be calculated from 
Eq.~(\ref{eq:NLSPtau}), where ${\cal B} = 1$ for our case where 
the NLSP is a slepton.  
The fractional uncertainty on the 
$\sqrt{F}$ measurement can be obtained adding in quadrature one 
fourth of the fractional error in $c\tau$ and five fourths of the 
fractional error on the slepton mass. 
In Fig.~\ref{fig:fractionalerror} (right side), we show the fractional 
error on the $\sqrt{F}$ measurement as a function of $\sqrt{F}$ for 
our three different assumptions on the $c\tau$ error. 
The uncertainty is dominated by $c\tau$ for the upper part of the 
$\sqrt{F}$ range, and grows quickly when approaching
the lower limit on $\sqrt{F}$. This is because very few sleptons 
survive and the statistical error on both $m_{\tilde \ell}$ and 
$c\tau$ gets very large. The error on $\sqrt{F}$ is better than 10\% 
for $1000 \ltap \sqrt{F} \ltap 4000$~TeV.

\section{Conclusions}
\label{sec:conc}

\noindent 
We have shown how the ATLAS detector at the LHC can be used to detect and 
measure the mass and lifetime of massive long-lived charged particles
produced in the framework of supersymmetric models with gauge-mediated SUSY
breaking, where a slepton is the NLSP and decays into a gravitino with a 
lifetime in the range 0.5~m $\ltap c\tau_{\rm NLSP} \ltap 1$~km.
At the LHC, a large amount of SUSY particles of this kind is generated,
while the SM background can be reduced at a very low level with 
appropriate cuts.

The lifetime measurement can be performed by a counting method with a good 
accuracy, allowing the computation of the value of the SUSY breaking scale 
$\sqrt{F}$.

The peculiar problems of the MSEP detection, related with the event
trigger and the time-of-flight measurement, have been discussed. The delicate
point of the possibility of measuring the specific ionisation by the
MDT's and by different sectors of the hadronic calorimeter has been 
addressed. We  stressed the importance that the specific ionisation 
measurements have for a clear MSEP identification.

\end{document}

%% file: SA_macros.tex
\newcommand{ \be  }{\begin{equation}}
\newcommand{ \ee  }{\end{equation}} 
\newcommand{ \bea }{\begin{eqnarray}}
\newcommand{ \eea }{\end{eqnarray}}
\newcommand{ \beas }{\begin{eqnarray*}}
\newcommand{ \eeas }{\end{eqnarray*}}

\newcommand{ \NPB    }[3]{Nucl. Phys.      {\bf B#1} (#2) #3}
\newcommand{ \PLB    }[3]{Phys. Lett. B    {\bf #1}  (#2) #3}
\newcommand{ \PLBold }[3]{Phys. Lett.      {\bf #1B} (#2) #3}
\newcommand{ \PRD    }[3]{Phys. Rev. D     {\bf #1}  (#2) #3}

\newcommand{ \PREP   }[3]{Phys. Rep.       {\bf #1}  (#2) #3}

\newcommand{ \EPJC   }[3]{Eur. Phys. J. C  {\bf #1}  (#2) #3}

\newcommand{ \centeron }[2]{{\setbox0=\hbox{#1}\setbox1=\hbox{#2}\ifdim
                             \wd1>\wd0\kern.5\wd1\kern-.5\wd0\fi \copy0
                             \kern-.5\wd0\kern-.5\wd1\copy1\ifdim\wd0>\wd1
                             \kern.5\wd0\kern-.5\wd1\fi}}
\newcommand{ \ltap }{\;\centeron{\raise.35ex\hbox{$<$}}
                     {\lower.65ex\hbox{$\sim$}}\;}
\newcommand{ \gtap }{\;\centeron{\raise.35ex\hbox{$>$}}
                     {\lower.65ex\hbox{$\sim$}}\;}

\newcommand{ \slashchar }[1]{\setbox0=\hbox{$#1$}   
   \dimen0=\wd0                                     
   \setbox1=\hbox{/} \dimen1=\wd1                   
   \ifdim\dimen0>\dimen1                            
      \rlap{\hbox to \dimen0{\hfil/\hfil}}          
      #1                                            
   \else                                            
      \rlap{\hbox to \dimen1{\hfil$#1$\hfil}}       
      /                                             
   \fi}                                             %

\def\G{ \tilde{G} }

\newcommand{ \tauu       }{ { \tilde \tau}_1 }


%% file: AMNPPRS.bbl
\vspace{0.7cm}

\begin{thebibliography}{10}

\bibitem{Errede}
  S.~Errede and S.~H.~H.~Tye, in Proc. of the 1984 Summer Study on the
  Design and Utilization of SSC, Snowmass, Colorado, 1984, 
  J.~Morfin and R.~Donaldson eds., pag.~175.

\bibitem{kaori}
  K.~Enqvist, K.~Mursula, M. Roos, \NPB{226}{1983}{121}. 

\bibitem{StevePrimer}
  S.~P.~Martin, ``A Supersymmetry Primer'', in ``Perspectives
  on Supersymmetry'', G.~L.~Kane ed., World Scientific 1998, 
  hep-ph/9709356, and references therein. 

\bibitem{oldGMSB}
  M.~Dine, W.~Fischler, M.~Srednicki, \NPB{189}{1981}{575};
  S.~Dimopoulos, S.~Raby, \NPB{192}{1981}{353};
  M.~Dine, W.~Fischler, \PLBold{110}{1982}{227};
  M.~Dine,  M.~Srednicki, \NPB{202}{1982}{238};
  M.~Dine, W.~Fischler, \NPB{204}{1982}{346};
  L.~Alvarez-Gaum\'e, M.~Claudson, M.~B.~Wise, \NPB{207}{1982}{96};
  C.~R.~Nappi, B.~A.~Ovrut, \PLBold{113}{1982}{175};
  S.~Dimopoulos, S.~Raby, \NPB{219}{1983}{479}.

\bibitem{newGMSB}
  M.~Dine, A.~E.~Nelson, \PRD{48}{1993}{1277};
  M.~Dine, A.~E.~Nelson, Y.~Shirman, \PRD{51}{1995}{1362};
  M.~Dine, A.~E.~Nelson, Y.~Nir, Y.~Shirman, \PRD{53}{1996}{2658}.

\bibitem{GR-GMSB}
  G.~F.~Giudice, R.~Rattazzi, ``Theories with 
  Gauge-Mediated Supersymmetry Breaking'', \PREP{322}{1999}{419-499, 501}.

\bibitem{GMSBmodels1}
  S.~Dimopoulos, S.~Thomas, J.~D.~Wells, \PRD{54}{1996}{3283};
  \NPB{488}{1997}{39}.

\bibitem{GMSBmodels2}
  J.~A.~Bagger, K.~Matchev, D.~M.~Pierce, R.~Zhang, 
  \PRD{55}{1997}{3188}. 

\bibitem{AKM-LEP2} 
  S.~Ambrosanio, G.~D.~Kribs, S.~P.~Martin, \PRD{56}{1997}{1761}. 

\bibitem{AB-LC}
  S.~Ambrosanio, G.~A.~Blair, \EPJC{12}{2000}{287--321}.

\bibitem{TDR}
  The ATLAS Collaboration, ``ATLAS Detector and Physics Performance Technical
  Design Report'', ATLAS TDR 15, CERN/LHCC/99-15, CERN Library, Geneva,
  Switzerland, 1999.

\bibitem{leandro}
  A.~Nisati, S.~Petrarca, G.~Salvini, Mod. Phys. Lett. {\bf A12} (1997) 2213;
  S.~Petrarca and G.~Salvini, ``Search for Stable Exotic Massive Particles at
  LHC by an Instrumented Air-Core Toroid (ASCOT type)'', Dip. di Fisica,
  Universit\`a di Roma ``La Sapienza'', Nota Interna n.~999 (1992).

\bibitem{drtata}
  M.~Drees, X.~Tata, \PLB{252}{1990}{695}.

\bibitem{femoroi}
  J.~L.~Feng, T.~Moroi, \PRD{58}{1998}{035001}.

\bibitem{marthom}
  S.~P.~Martin, J.~D.~Wells, \PRD{59}{1999}{035008}.

\bibitem{Tata}
  M.~Drees and X.~Tata, \PLB{252}{1990}{695}.

\bibitem{frontend} 
  The ATLAS Collaboration, ``Muon Spectrometer Technical Design
  Report'', CERN/LHCC/97-22, CERN Library, Geneva, Switzerland, 1997. 

\bibitem{leandro1}
  A.~Nisati, ``Preliminary Timing Studies of the Barrel Muon Trigger 
  System'', ATLAS Internal Note ATL-DAQ-98-083, CERN Library, Geneva,
  Switzerland, 1998.

\bibitem{gpar}
  G.~Polesello, A.~Rimoldi, ``Reconstruction of Quasi-stable Charged
  Sleptons in the ATLAS Muon Spectrometer'', ATLAS Internal Note 
  ATL-MUON-99-06, CERN Library, Geneva, Switzerland, 1999.

\bibitem{Fayet}
  P.~Fayet, \PLBold{70}{1977}{461}; \PLBold{86}{1979}{272};
  \PLB{175}{1986}{471} and in ``Unification of the fundamental 
  particle interactions", S.~Ferrara, J.~Ellis,   
  P.~van Nieuwenhuizen eds. (Plenum, New York, 1980), p.~587.

\bibitem{SUSYFIRE}
  An updated, generalised and {\tt Fortran}-linked version of the 
  program used in Ref.~\cite{AKM-LEP2}. It generates minimal and 
  non-minimal GMSB and SUGRA models.
  For inquiries about this software package, please send e-mail to  
  {\tt ambros@mail.cern.ch}. 

\bibitem{AMPPR1}
  S.~Ambrosanio, B.~Mele, S.~Petrarca, G.~Polesello, A.~Rimoldi,
  in ``Aspects of GMSB Phenomenology at TeV Colliders'', in hep/ph-0002191
  and hep/ph-0005142, Summary Report of the SUSY Working Group, 
  Workshop on Physics at TeV Colliders, Les Houches, France, 7-18 June 1999.
  
\bibitem{AMPPR2}
  S.~Ambrosanio, B.~Mele, S.~Petrarca, G.~Polesello, A.~Rimoldi, 
  ``Measuring the SUSY Breaking Scale at the LHC in the Slepton 
  NLSP Scenario of GMSB Models'', \\ hep-ph/0010081, submitted to
  {\it The Journal of High-Energy Physics}. 

\end{thebibliography}
